\documentclass[11pt]{article}
\pdfoutput=1 
\usepackage{amssymb}
\usepackage{amsmath}
\usepackage{amstext}
\usepackage{graphicx,epsfig}
\usepackage{epsfig}
\usepackage{verbatim} 
\usepackage{fancybox}
\usepackage{color}
\usepackage{ulem}
\usepackage{enumitem}
\usepackage{youngtab}
\usepackage{bbm}
\usepackage{parskip}
\usepackage[numbers,sort&compress]{natbib}
\usepackage{ytableau}
\usepackage[utf8]{inputenc}

\usepackage{caption}
\usepackage{subcaption}

\newcommand{\Comment}[1]{{}}
\definecolor{darkblue}{rgb}{0.15,0.35,0.55}
\definecolor{reddish}{rgb}{0.65, 0.2, 0.2}
\usepackage[linktocpage=true]{hyperref}
\hypersetup{
colorlinks=true,
citecolor=darkblue,
linkcolor=reddish,
urlcolor=darkblue,
pdfauthor={},
pdftitle={},
pdfsubject={}
}

\definecolor{green3}{RGB}{44, 160, 44}

\newcommand{\be}{\begin{equation}}
\newcommand{\ee}{\end{equation}}
\newcommand{\bea}{\begin{eqnarray}}
\newcommand{\eea}{\end{eqnarray}}
\newcommand{\beas}{\begin{eqnarray*}}
\newcommand{\eeas}{\end{eqnarray*}}

\definecolor{darkred}{rgb}{0.7,0.3,0.3}
\definecolor{darkgreen}{rgb}{0.2,0.7,0.3}
\definecolor{lightgreen}{rgb}{.816,.94,.753}
\definecolor{greyish}{rgb}{.8,.8,.8}
\definecolor{darkblue2}{rgb}{0.3,0.4,0.9}
\usepackage{pifont}
\usepackage{xcolor,colortbl}

\def\({\left(}
\def\){\right)}

\def\gsim{ \lower .75ex \hbox{$\sim$} \llap{\raise .27ex \hbox{$>$}} }
\def\lsim{ \lower .75ex \hbox{$\sim$} \llap{\raise .27ex \hbox{$<$}} }

\def\xyma{\xymatrix@M.7em}
\def\xymas{\xymatrix@M.1em}

\title{}
\author{}

\numberwithin{equation}{section}

\begin{document}

\renewcommand{\thefootnote}{\fnsymbol{footnote}}
~

\begin{center}
{\huge \bf Casimir Tests of Scalar-Tensor Theories}\\
\end{center} 

\vspace{1truecm}
\thispagestyle{empty}
\centerline{\Large 
Philippe Brax,${}^{\rm a,b}$\footnote{\href{mailto:philippe.brax@ipht.fr}{\texttt{philippe.brax@ipht.fr}}}
Anne-Christine Davis,${}^{\rm c}$\footnote{\href{mailto:acd@damtp.cam.ac.uk} {\texttt{acd@damtp.cam.ac.uk}}}
and Benjamin Elder${}^{\rm d}$\footnote{\href{mailto:bcelder@hawaii.edu} {\texttt{bcelder@hawaii.edu}}}

}

\vspace{.5cm}
 
\centerline{{\it ${}^{\rm a}$Institut de Physique Th\'eorique,}}
 \centerline{{\it Universit\'e Paris-Saclay, CEA, CNRS, F-91191 Gif-sur-Yvette, France}} 
 \vspace{.25cm}

\centerline{{\it ${}^{\rm b}$ CERN, Theoretical Physics Department, Geneva, Switzerland.}}
 \vspace{.25cm}
 
\centerline{{\it ${}^{\rm c}$DAMTP, Centre for Mathematical Sciences,}}
 \centerline{{\it University of Cambridge, Cambridge CB3 0WA, United Kingdom}} 
 \vspace{.25cm}
 
 
  \centerline{{\it ${}^{\rm d}$Department of Physics and Astronomy,}}
 \centerline{{\it University of Hawai'i, Honolulu 96822, United States of America}} 

 \vspace{1cm}
\begin{abstract}
\noindent
We compute bounds and forecasts on screened modified gravity theories, specialising to the chameleon model in Casimir force  experiments. In particular,
we investigate the classical interaction between a plate and sphere subject to a screened interaction of the chameleon type. We compare numerical simulations of the field profile and the classical pressure exerted on the sphere to analytical approximations for these non-linear field theories. In particular, we focus on the proximity force approximation (PFA) and show that, within the range of sphere sizes $R$ and plate-sphere distance $D$ simulated numerically, the PFA does not reproduce the numerical results. This differs from the case of linear field theories such as Newtonian gravity and a Yukawa model where the PFA coincides with the exact results. We show that for chameleon theories, the  screening factor approximation (SFA) whereby the sphere is modelled as a screened sphere embedded in the external field due to the plates,  fares better and can be used in the regime $D\gtrsim R$ to extract constraints and forecasts from existing and forthcoming data. In particular, we forecast that future Casimir experiments would corroborate the closing of the parameter space for the simplest of chameleon models at the dark energy scale. 

\end{abstract}

\newpage

\setcounter{tocdepth}{2}
\renewcommand*{\thefootnote}{\arabic{footnote}}
\setcounter{footnote}{0}

\newpage
\section{Introduction}
Modified gravity theories whereby the Einstein-Hilbert action is supplemented with an additional scalar field have been extensively developed over the years. Such theories open up new testing grounds for General Relativity on a range of scales not previously accessible. They could also explain puzzles in standard model physics, like the anomalous magnetic moment of the muon~\cite{Brax:2021owd} or the observed acceleration of the expansion of the Universe~\cite{CANTATA:2021ktz}. However, the existence of such scalar fields is severely constrained by fifth force experiments. In modern developments the scalar force is {\it screened} such that Solar System constraints are evaded.

Such modifications of gravity by screened scalar interactions are now classified and have been tested from the cosmological scales down to the laboratory~\cite{Joyce:2014kja, CANTATA:2021ktz, Brax:2021wcv}. They appear in three types and are characterised by lower bounds on the Newtonian potential or  its first and second spatial gradients. Non-linear scalar field theories where screening appears when the Newtonian potential of test objects is large enough are chameleon screened~\cite{Khoury:2003aq} and can be tested in the laboratory thanks to a large variety of experiments, e.g.  levitating spheres, Q-bouncing neutrons, atomic interferometry~\cite{Burrage:2014oza,Hamilton:2015zga,Elder:2016yxm,Burrage:2016rkv, Jaffe:2016fsh, Sabulsky:2018jma}, torsion balance experiments~\cite{Upadhye:2012qu,Upadhye:2012rc} and Casimir effect measurements~\cite{Brax:2007vm,Brax:2018grq}.  (See~\cite{Burrage:2016bwy, CANTATA:2021ktz, Brax:2021wcv} for reviews of these bounds.) In this paper we concentrate on Casimir tests. The Casmir interaction is a prominent quantum interaction in the micrometre range and is exerted between two metallic plates. The original calculation by Casimir does not take into account temperature effects and the finite conductivity of real material. Reviews on the  modification to the Casimir interaction from temperature, finite conductivity and surface roughness effects can be found in \cite{Klimchitskaya:2009cw}. Here we will consider a regime where distances between objects is large enough that the Casimir pressure due to the quantum fluctuations of the photon can be small enough to allow for other interactions to feature. Of course a nearly massless scalar field would immediately lead to an enhancement of the photon Casimir interaction by a factor of $3/2$ and would be experimentally ruled out. For a thorough discussion of the quantum Casimir effect for light scalar fields, see \cite{Brax:2018grq}.  We will be interested here in a more subtle effect when the Compton wavelength  of the scalar field in the experimental environment is 
smaller than the typical distance between the test objects. In this regime, the quantum fluctuations of the scalar field are sufficiently suppressed to be neglected. This is particularly relevant for screened models of the chameleon type~\cite{Khoury:2003aq} where the fields do not penetrate in the test objects and a classical interaction can be generated between them. This effect is the classical equivalent of an electrostatic interaction for photons.

The classical interaction generated by chameleons between screened objects has already been exploited and useful bounds on the parameter space have ensued~\cite{Brax:2007vm}. In this paper, we come back to this issue in order to ascertain whether the hypotheses used previously to tackle the calculation of the classical pressure are indeed valid. No problems exist in the case of a plane-plane situation where the field profile and the pressure on the plates can be calculated exactly thanks to the planar symmetry of the configuration. Unfortunately, the plate-plate configuration is not the experimentally favoured one as maintaining two planes exactly parallel in an experimental setup is particularly difficult. The plane-sphere geometry is more appropriate but also less amenable to an exact analytical treatment. In the case of Newtonian gravity, the exact force can be retrieved using the ``proximity force approximation'' (PFA, also called the Derjaguin approximation)~\cite{derjaguin_1934} whereby each spherical element at the surface of the sphere can be approximated by its tangent plane and the interaction between the sphere and the plate can be obtained as the sum of all the elemental pressures on the tangent planes exerted by the infinitely large plate. More precisely each elemental plane bounds a small cylinder which is attracted by the infinite plane. The sum of forces on each of the infinitesimal cylinders coincide with the Newtonian pressure between the plane and the sphere. The same method can be applied to a Yukawa interaction involving a massive scalar field linearly coupled to matter~\cite{Decca:2009fg}, and has been refined to describe electromagnetic forces~\cite{Fosco:2012jc}. In previous work~\cite{Brax:2007vm}, the proximity approximation was also extended to chameleon theories. In this paper, we investigate the validity of this extension by comparing numerical simulations when $D\gtrsim R$, where $D$ is the sphere-plane distance and $R$ the radius of the sphere, to the proximity approximation. We show that the PFA fails to reproduce the numerical results. Surprisingly, another approximation where the sphere is treated as a screened sphere embedded in the external field created by the plate more accurately reproduces the numerical results. We then use this approximation to forecast what future Casimir experiments could measure and show that for the most popular of chameleon models, the $n=1$ inverse power law chameleon, the future Casimir measurements would close part of the chameleon parameter space. This part of the parameter has recently been closed experimentally~\cite{2022NatPh..18.1181Y} and Casimir experiments would confirm this in a different context. Moreover, we discuss how improving the sensitivity of future Casimir experiments would further this result. We extend our results to the n=2 chameleon using our numerical data and to arbitrary n chameleons using the SFA approximation. This enables us to put bounds on the parameter space for this class of theories. This work is akin to our previous paper where forecasts were made for future Casimir experiments for the symmetron model of screened modified gravity~\cite{Elder:2019yyp}, although our findings differ significantly.  In the case of the symmetron, distinct regimes were identified in which the PFA was accurate.  In the case of the chameleon, we will see that we are unable to confirm the accuracy of this approximation method, even within the regime where one might expect it to work well.  We use the same experimental prescription for these forecasts.

The paper is arranged as follows. In Sec.~\ref{sec:ST}, we discuss the calculation of the plate-sphere interaction in the case of Newtonian and Yukawa interactions. We show how the Proximity Force Approximation reproduces the exact result which can be calculated exactly. In Sec.~\ref{sec:cham} we introduce the chameleon models and consider the Casimir interaction between two plates and between a plate and a sphere. We introduce another approximation, the Screening Factor Approximation (SFA) whereby the screened sphere is embedded in the field profile generated by the planar plate. In Sec.~\ref{sec:num} we give the results of numerical simulations for the field profile and the force when the radius of the sphere and its distance to the plate is varied. We show that the SFA reproduces the numerical results much better than the PFA.  Finally, in Sec.~\ref{sec:exp}, we use current experimental specifications to forecast the reach of future Casimir experiments for chameleons. We also investigate how sensitive these experiments should be to close the parameter space almost completely. Finally we conclude in Sec.~\ref{sec:conclusions}.

\section{Sphere-plate interactions in scalar-tensor theories}
\label{sec:ST}

Casimir-like experiments are sensitive to new physics like the force mediated by a screened scalar field. This is because such a test involves high precision and a very high vacuum between dense objects. In this context, models which pass the Solar System tests of gravity can be probed in the laboratory. 

We are interested  in computing the classical force between an infinite plate and a finite sphere in scalar-tensor theories, which may be written in the following form:
\begin{equation}
    S = \int d^4 x \sqrt{-g} \left( \frac{M_\mathrm{Pl}^2}{2} R - \frac{1}{2} (\partial \phi)^2 - V(\phi) \right) + S_\mathrm{matter}[A^2(\phi) g_{\mu_\nu}; \psi]~.
\end{equation}

The scalar force on a pointlike object of mass $m_\mathrm{obj}$ in this theory is
\begin{equation}
    \vec F = - m_\mathrm{obj}\frac{\mathrm{d} \ln A}{\mathrm{d} \phi} \vec \nabla \phi~,
\end{equation}
while the force on an extended object in an external field is given by
\begin{equation}
    \vec F = - m_\mathrm{obj} \lambda_\mathrm{obj} \frac{\mathrm{d} \ln A}{\mathrm{d} \phi} \vec \nabla \phi_\mathrm{ext}(\vec x)~.
\end{equation}
The scalar charge carried by the object is $Q_\mathrm{obj} = \lambda_\mathrm{obj} m_\mathrm{obj}$, where $\lambda_\mathrm{obj}$ is referred to as the ``screening factor'' and varies between zero (for large, dense objects) and 1 (for small, light objects).

Our previous paper \cite{Elder:2019yyp} carefully worked out the classical force between an infinite plate and a finite sphere in the case of symmetron models. Different regimes can be discriminated depending on the mass of scalar field in vacuum, i.e. the mass the scalar field would have in the experimental vacuum in the absence of boundary plates\footnote{The actual mass of the scalar field becomes space-dependent between the plates as the field profile is non-trivial. We  assume that the typical mass of the scalar between the plates is larger than $1/D$. This suppresses the quantum interaction mediated by the scalar field which would be ruled out if present.}.  This suggests the following generalization for obtaining the force in most scalar-tensor theories:
\begin{itemize}
    \item When $R \ll m_\mathrm{vac}^{-1}$, write down the exact field profile $\phi_\mathrm{plate}$ sourced by an infinite plate.  The force is then the one due to the external field generated by the plate acting on the sphere simply  embedded in the external profile. We will refer to this approximation as the ``screening factor approximation'' (SFA)
    \begin{equation}
        F = - M_\mathrm{sphere} \lambda_\mathrm{sphere} A,_\phi \vec \nabla \phi_\mathrm{plate}~.
    \end{equation}
    where the gradient is evaluated at the centre of the sphere.

    \item When $R \gg m_\mathrm{vac}^{-1}$, solve for the exact pressure between two parallel plates, $P_\mathrm{parallel}(L)$, which is in general a function of their separation $L$.  Then we can approximate the force on the sphere as
    \begin{equation}
        F = - 2 \pi R^2 \int_{\pi/2}^\pi d \theta \sin\theta \cos \theta ~P_\mathrm{parallel}( D + R + R\cos \theta)~.
    \end{equation}
    The will be referred to as the ``proximity force approximation'' (PFA).
    
    \item When $R \approx m_\mathrm{vac}^{-1}$, the problem must be solved numerically.  Alternatively, the results from the previous two limits may be interpolated to approximate the force in this regime.
\end{itemize}

In the following sections we will apply this prescription to scalar-tensor theories, in particular the inverse power law chameleon, which are already critically constrained by experiments such as atomic interferometry. We will find that the PFA is not reliable as an approximation and that in the long distance regime $D\gtrsim R$ the SFA is much more accurate. 
Before focusing on non-linear scalar-tensor theories where screening takes place, we will review the status of the sphere-plate interaction in Newtonian gravity and Yukawa theories. We will emphasize the role of the PFA in this context. 

\vspace{.15cm}
\noindent
{\bf Conventions:}
We are adopting units where $c = \hbar = 1$, and define $M_\mathrm{Pl} = (8 \pi G)^{-1/2}$.  We work in the mostly-plus metric convention, and derivatives of fields are denoted with a subscript, e.g. $\frac{d}{d \phi} A(\phi) = A,_\phi$~.

\subsection{Proximity Force Approximation in Newtonian Gravity}
As a warm-up we will first develop the PFA in Newtonian gravity.  This will be helpful because, in addition to the solutions being simpler, there are several remarkable properties that will allow us to obtain the exact force between the sphere and plate.  Furthermore, there is no screening in this theory, this case is considerably simpler and will allow us to focus on the important differences for screened theories later on.

The Newtonian potential $\phi$ is given by Poisson's equation:
\begin{equation}
    \vec \nabla^2 \phi = 4 \pi G \rho~.
    \label{poisson}
\end{equation}
The force on a test particle of mass $m$ is $\vec F = - m \vec \nabla \phi$.  Since we only measure the force, it is often helpful to work with the (Newtonian) gravitational field, $\vec g \equiv - \vec \nabla \phi$, so that the force on a test particle is $\vec F = m \vec g$, and the equation of motion is
\begin{equation}
    \vec \nabla \cdot \vec g = - 4 \pi G \rho~.
    \label{grav-field-eqn}
\end{equation}
First we need to solve for the gravitational field $\vec g$ around an infinite plate of uniform density.  As  there is no screening, we must introduce a plate thickness $T$ and density $\rho_\mathrm{plate}$.  Although we could integrate Eq.~\eqref{poisson} directly, a shortcut is to use the divergence theorem, which works with Eq.~\eqref{grav-field-eqn}:
\begin{equation}
    \int_V (\vec \nabla \cdot \vec g) dV = \int_S (\vec g \cdot \hat n) dS~.
\end{equation}
This allows us to turn a volume integral over $V$ into a surface integral on the boundary $S$ with outward-pointing normal vector $\hat n$.  Applying this theorem to Eq.~\eqref{poisson}, we find
\begin{equation}
    \int_S (\vec g \cdot \hat n) dS = - 4 \pi G m~,
\end{equation}
where $m$ is the mass enclosed within the surface S~.

It is now straightforward to solve for the field around a plate.  Draw a cube with sides of length $l > T$ centered on the plate.  By symmetry, $\vec g$ is perpendicular to the sides of the box, so we just have to integrate over the faces that are parallel to the plate.  The field is uniform, so we have
\begin{equation}
    2 |\vec g| l^2 = - 4 \pi G \rho_\mathrm{plate} T l^2~.
\end{equation}
Isolating $\vec g$, and recalling that the contribution to the surface integral is positive when the gravitational field is parallel to the outward-facing normal vector $\hat n$, we have
\begin{equation}
    \vec g = - 2 \pi G \rho_\mathrm{plate} T \hat x \quad \mathrm{for} \quad \vec e_x > 0
\end{equation}
where $\vec e_x$ is the unit vector along the $x$ axis, pointing towards positive $x$ values. 
Notice that this field is uniform, so an object of mass $m$ and arbitrary shape will move like a test particle located at the object's center of mass.

It is  now trivial to solve for the exact force between the sphere and the plate:
\begin{equation}
    \vec F = - 2 \pi G \rho_\mathrm{plate} T m_\mathrm{sphere} \vec e_x.
    \label{force-sphere-plate-poisson}
\end{equation}

Now let us  think about how we might reproduce Eq.~\eqref{force-sphere-plate-poisson} with the PFA.  There is no need to do this for the present case since we already have the exact result, but it will give us some ideas of how to generalize the PFA to deal with objects of finite thickness in more complicated theories.

The starting point is to solve for the pressure between two parallel plates, of density $\rho_i$ and thickness $T_i$, separated by a distance $L$~.  The force on a section of plate 2, enclosed within a box with sides of length $l > T$ is
\begin{equation}
    \vec F = - 2 \pi G \rho_1 T_1 \rho_2 T_2 l^2 \vec e_x~,
\end{equation}
where the minus sign indicates an attractive force.  The pressure between the plates is then
\begin{equation}
    P_x = - 2 \pi G \rho_1 T_1 \rho_2 T_2~.
\end{equation}
This expression is independent of $L$ thanks to the uniform gravitational field source by the plates.

\begin{figure}[t]
\centering
\includegraphics[width=0.5\textwidth]{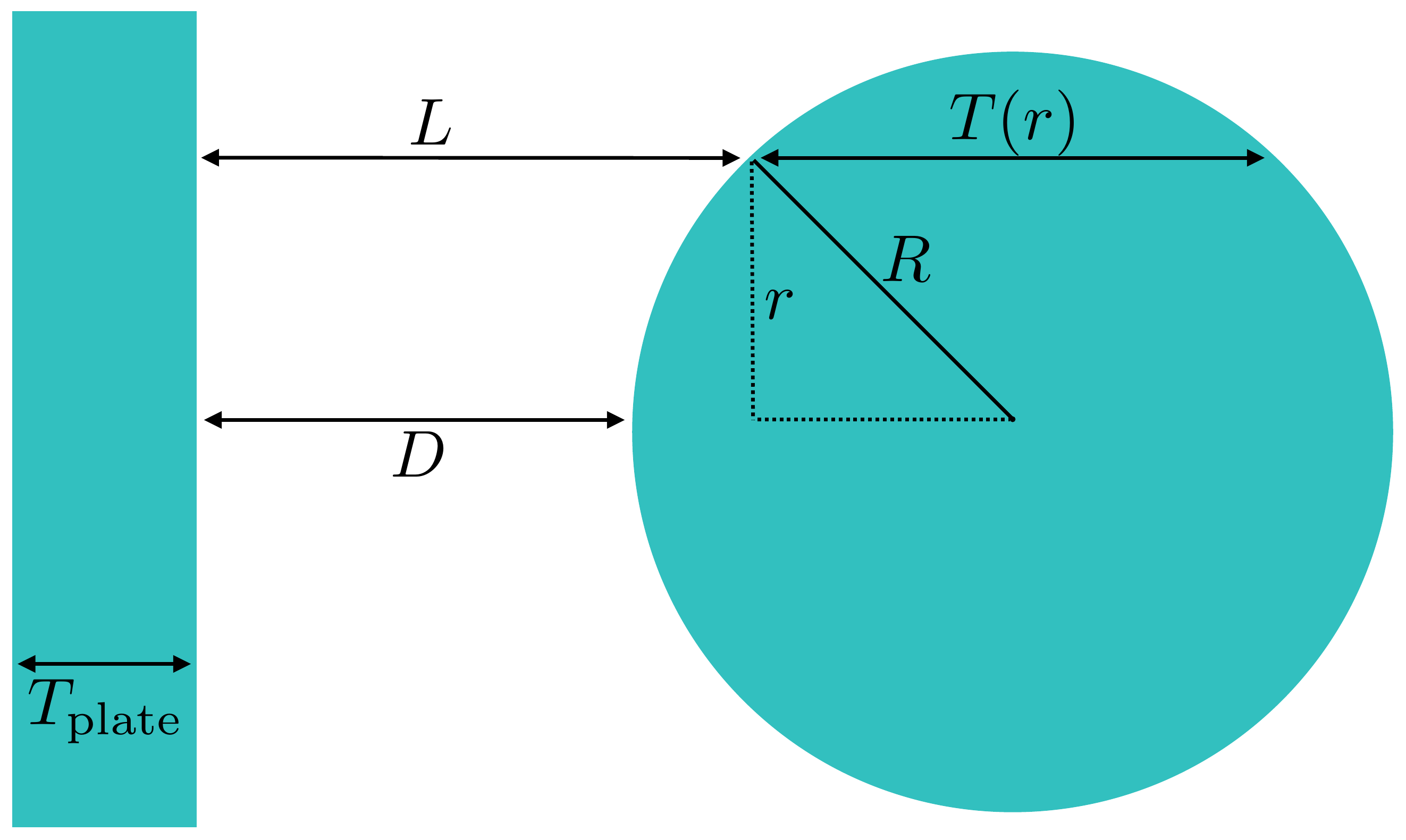}
\caption{\small Sphere-plate system, where the thickness of the second plate in the PFA is specified as $T(r)$.}
\label{fig-sphere-thickness}
\end{figure}

Now let us use this result in the PFA to approximate the force between the sphere and the plate.  We immediately have a challenge, because our earlier formulation of the PFA accounts for the plate separation $L$, but makes no reference to the thicknesses of the materials.  One possible modification suggests itself: when considering the force on an infinitesimal patch $d A$ on the surface of the sphere, we will use the pressure that corresponds to a plate of thickness $T_2$ that is equal to the thickness of of the sphere along a line perpendicular to the plate, which passes through $dA$, as shown in Fig.~\ref{fig-sphere-thickness}.

The advantage of this prescription is that it will give us a final result that is exactly correct in this case.  Noting that the force on $dA$ is the same for any other patch located on a ring that is parallel to the plane, we see that we are simply breaking the sphere up into cylinders of infinitesimal thickness.  Each cylinder has a mass
\begin{equation}
    m(r) = \rho_\mathrm{sphere} 2 \pi r dr T(r)~,
\end{equation}
where $r$ is the radius of the cylinder and T(r) is the height.  We can treat this as a test object of mass $m(r)$ located at the center of mass, and the total force on the sphere is obtained by integrating all cylinder radii $0 < r < R$~.  By geometry, $T(r) = 2\sqrt{R^2 - r^2}$, and the net force is
\begin{align} \nonumber
    F &= - 2 \pi G T_\mathrm{plate} \rho_\mathrm{plate} \int_0^R 4 \pi r \sqrt{R^2 - r^2} dr \rho_\mathrm{sphere}~, \\
    &= - 2 \pi G T_\mathrm{plate} \rho_\mathrm{plate}  m_\mathrm{sphere}~,
\end{align}
which precisely agrees with the exact result.
This version of the PFA describes the Newtonian force between a sphere and a plate exactly. 

\subsection{Yukawa interactions}
As a next step, we consider a Yukawa theory of a massive scalar field coupled linearly to matter:
\begin{equation}
    {\cal L} = - \frac{1}{2} (\partial \phi)^2 - \frac{1}{2} m^2 \phi^2 - \frac{\phi}{M} \rho~.
\end{equation}
First we must calculate how the field responds to a sphere in isolation. The static equation of motion is
\begin{equation}
    \vec \nabla^2 \phi = m^2 \phi + \frac{\rho}{M}~.
\end{equation}
We can bring this into a more familiar form by making the variable transformation
\begin{equation}
    y \equiv \phi + \frac{\rho}{m^2 M}~,
\end{equation}
so that the equation of motion is
\begin{equation}
    \vec \nabla^2 y = m^2 y~,
\end{equation}
which has the spherically symmetric solution
\begin{equation}
    y = A\frac{e^{m r}}{r} + B \frac{e^{- m r}}{r}~.
\end{equation}
Transforming back to our original variable $\phi$, we find
\begin{equation}
    \phi = A \frac{e^{m r}}{r} + B \frac{e^{- m r}}{r} - \frac{\rho}{m^2 M}~.
\end{equation}
The density is a piecewise function of $r$~,
\begin{equation}
    \rho(r) = \begin{cases}
    \rho_\mathrm{in} & \mathrm{for} ~ r < R~, \\
    \rho_\mathrm{out} &  \mathrm{for} ~ r > R~.
    \end{cases}
\end{equation}
We solve the equation of motion in a piecewise manner, with four undetermined constants of integration:
\begin{equation}
    \phi = \begin{cases}
    \phi_\mathrm{in} \equiv A_\mathrm{in} \frac{e^{m r}}{r} + B_\mathrm{in} \frac{e^{- m r}}{r} - \frac{\rho_\mathrm{in}}{m^2 M} & \mathrm{for} ~ r < R~, \\
    \phi_\mathrm{out} \equiv A_\mathrm{out} \frac{e^{m r}}{r} + B_\mathrm{out} \frac{e^{- m r}}{r} - \frac{\rho_\mathrm{out}}{m^2 M} &  \mathrm{for} ~ r > R~.
    \end{cases}
\end{equation}
Our boundary conditions are the following: $\phi(r \to \infty) = - \frac{\rho_\mathrm{out}}{m^2 M}$, $\phi'(0) = 0$~, $\phi_\mathrm{in}(R) = \phi_\mathrm{out}(R)$~, $\phi_\mathrm{in}'(R) = \phi_\mathrm{out}'(R)$.
The first condition demands $A_\mathrm{out} = 0$, and the second condition enforces  $B_\mathrm{in} = - A_\mathrm{in}$ so we have
\begin{equation}
    \phi_\mathrm{in} = A_\mathrm{in} \frac{\sinh m r}{r} - \frac{\rho_\mathrm{in}}{m^2 M}~.
\end{equation}
We will henceforth denote $A \equiv A_\mathrm{in}$ and $B \equiv B_\mathrm{out}$~.  The matching conditions at $R$ gives a set of equations:
\begin{align} \nonumber
    A \frac{\sinh m R}{R} - \frac{\rho_\mathrm{in} - \rho_\mathrm{out}}{m^2 M} &= B \frac{e^{- m R}}{R}~, \\
    A \left( \cosh m R - \frac{\sinh m R}{m R} \right) &= - B \left( \frac{1 + mR}{mR} \right) e^{- m R}~.
\end{align}
We solve these for $B$ only, since we are only interested in  the external solution, finding
\begin{equation}
    B = - \frac{\rho_\mathrm{in} - \rho_\mathrm{out}}{m^3 M} \left( \frac{m R - \tanh mR}{1 + \tanh mR} \right)e^{m R}~,
\end{equation}
so the external field profile is now
\begin{equation}
    \phi(r > R) = - \frac{\rho_\mathrm{in} - \rho_\mathrm{out}}{m^3 M} \left( \frac{m R - \tanh mR}{1 + \tanh mR} \right) \frac{e^{-m(r -R)}}{r} - \frac{\rho_\mathrm{out}}{m^2 M}~,
    \label{full-external-soln}
\end{equation}
Let us now rearrange this in a way that lets us identify the scalar charge, as well as a screening factor.  We take the point-particle limit, $m R \ll 1$, and at leading order we have
\begin{equation}
    \lim_{mR \to 0} \phi(r) = - \frac{m_\mathrm{sph}}{4 \pi M} \frac{e^{-m(r -R)}}{r} - \frac{\rho_\mathrm{out}}{m^2 M}~,
\end{equation}
where $m_\mathrm{sph} = 4 \pi \rho_\mathrm{in} R^3 / 3$ is the mass of the sphere.
We thus define the screening factor in the following way:
\begin{equation}
    \phi(r) \equiv - \frac{\lambda_\mathrm{sph} m_\mathrm{sph}}{4 \pi M} \frac{e^{- m(r - R)}}{r} - \frac{\rho_\mathrm{out}}{m^2 M}~.
    \label{screening-factor-def}
\end{equation}
The screening factor $\lambda_\mathrm{sph}$ encodes the finite-size effects of the sphere.  Matching Eqs.~\eqref{full-external-soln} and \eqref{screening-factor-def} leads us to conclude
\begin{equation}
    \lambda_\mathrm{sph} = \frac{3}{(m R)^3} \left( 1 - \frac{\rho_\mathrm{out}}{\rho_\mathrm{in}} \right) \left( \frac{m R - \tanh mR}{1 + \tanh mR} \right)~,
\end{equation}
and the scalar charge is  $Q_\mathrm{sph} \equiv \lambda_\mathrm{sph} m_\mathrm{sph}$~.

Now that we have the screening factor, we can compute the force on the sphere in the following way:
\begin{equation}
    \vec F = - \frac{\lambda_\mathrm{sph} m_\mathrm{sph}}{M} \vec \nabla \phi_{\rm ext}~.
    \label{force-screening}
\end{equation}
If we use this to compute the force between two unscreened test particles, then we find an interaction potential
\begin{equation}
    V(r) = - \frac{m_1 m_2}{4 \pi M^2} \frac{e^{- m r}}{r}~,
\end{equation}
which allows us to match our notation to  Eq.~(10) in \cite{Decca:2009fg}~.  This will enable us to compare our results to those previously obtained.  For reference, that mapping is
\begin{equation}
    \frac{1}{4 \pi M^2} = G \alpha~, \quad \quad m = \frac{1}{\lambda}~,
\end{equation}
where the notation of the present work is on the left hand side of the equations.

Now we use the screening factor to compute the force between an infinite plate of finite width ($2 W$) and a sphere of radius $R$.  The nearest distance between the surface of the sphere and the plate is $D$.   We must first solve for the field configuration around the plate in isolation.  Due to the planar symmetry, the general solution to the equation of motion is
\begin{equation}
    \phi = A e^{m x} + B e^{- m x} - \frac{\rho}{m^2 M}~.
\end{equation}
We solve in a piecewise manner again inside and outside the plate, with $x = 0$ located in the middle of a plate of width $2 D$.  Requiring $\phi$ to be finite at $x = \infty$ and regular at the origin leads us to
\begin{equation}
    \phi(x > 0) = \begin{cases}
    A \cosh m x - \frac{\rho_\mathrm{in}}{m^2 M} & \mathrm{for~} x < W~, \\
    B e^{- m x} - \frac{\rho_\mathrm{out}}{m^2 M} & \mathrm{for~} x > W~.
    \end{cases}
\end{equation}
This profile is symmetric about $x = 0$.
Matching the field and its first derivative at $x = W$ gives the matching conditions
\begin{align} \nonumber
    A \cosh m W - \frac{\rho_\mathrm{in} - \rho_\mathrm{out}}{m^2 M} &= B e^{- m W}~,\\
    A \sinh m W &= - B e^{- m W}~.
\end{align}
Adding these equations together, we find
\begin{equation}
    A = \frac{\rho_\mathrm{in} - \rho_\mathrm{out}}{m^2 M} e^{- m W}
\end{equation}
and then plugging this into the second matching condition yields
\begin{equation}
    B = - \frac{\rho_\mathrm{in} - \rho_\mathrm{out}}{m^2 M} \sinh m W~.
\end{equation}
Putting everything together, we find that the field around a plate of total width $W_1$  is 
\begin{equation}
    \phi_\mathrm{plt}(x) = - \frac{\rho_\mathrm{in} - \rho_\mathrm{out}}{m^2 M} \sinh \frac{m W_1}{2}  e^{- m x} - \frac{\rho_\mathrm{out}}{m^2 M}~,
\end{equation}
where $x$ is the distance from the center of the plate.  Finally, we can combine this result with Eq.~\eqref{force-screening} to obtain the force between a plate of width $W_1$ and a sphere of radius $R$ separated by a distance $D$:
\begin{align} \nonumber
    F = &- \frac{3}{(m R)^3} \left( 1 - \frac{\rho_\mathrm{out}}{\rho_\mathrm{sph}} \right) \left( \frac{m R - \tanh mR}{1 + \tanh mR} \right) \left( \frac 4 3 \pi \rho_\mathrm{sph} R^3 \right) \\ 
    &\times \frac{\rho_\mathrm{plt} - \rho_\mathrm{out}}{m^2 M} \sinh \frac{mW_1}{2} \left(\frac{m}{M} \right) e^{- m (D + R + W_1/2)}
\end{align}
We will  henceforth assume $\rho_\mathrm{out} = 0$, and simplifying the expression  we find
\begin{equation}
    F = - \frac{2 \pi \rho_\mathrm{sph} \rho_\mathrm{plt}}{m^4 M^2} \left( \frac{m R - \tanh mR}{1 + \tanh mR} \right) e^{- m R} \left( 1 - e^{- m W_1} \right) e^{- m D}~.
    \label{force-screened-sphere-2}
\end{equation}
Recall that the screening factor calculation is only valid when the background field is approximately linear over the extent of the sphere.  Applying Eq.~\eqref{screening-validity}, we find the condition for validity of this result is
\begin{equation}
    \phi'_\mathrm{plt}(x) \gg \phi''_\mathrm{plt}(x) R~,
\end{equation}
which is satisfied when $m R \ll 1$.  In this limit, our result to leading order is
\begin{equation}
    F \approx \frac{2 \pi \rho_\mathrm{sph} \rho_\mathrm{plt}}{3 m^4 M^2} (m R)^3 \left( 1 - e^{- m W_1} \right) e^{- m D} + O(m R)^4~.
\end{equation}

The proximity calculation  can be carried in a similar fashion to the one for Newtonian gravity. First of all, the field inside a plate of density $\rho_2$ and width $W_2$ at a distance $2d$ from another plate of density $\rho_1$ and width $W_1$ is given by
\begin{equation}
\phi(x) = -\frac{\rho_2}{m^2 M}- (1-e^{-mW_1}) \frac{\rho_1}{2m^2 M} e^{m(x+d)}+ \frac{\rho_2}{2m^2 M}(e^{m(d-x)}+ e^{m(x-d-W_2)}).
\end{equation}
where $x$ is between $d$ and $d+W_2$.
The pressure on the plate is given by
\begin{equation}  
P= -\frac{\rho_2}{M}\int_{d}^{d+W_2}\partial_x \phi= \frac{\rho_2}{M}(\phi(d)-\phi(d+W_2)
\end{equation}
which gives
\begin{equation}
    P= -\frac{\rho_1\rho_2}{2m^2 M}(1-e^{-mW_1})(1-e^{-mW_2})e^{-2md}.
\end{equation}
With this result we can calculate the force due to a plate on a sphere as
\begin{equation}
    F_{\rm PFA}=- \frac{\pi\rho_\mathrm{sph} \rho_\mathrm{plt}}{m^2 M}(1-e^{-mW_1})\int_0^R dr r e^{D+R-\sqrt{R^2-r^2}}(1-e^{-2m\sqrt{R^2-r^2}})
\end{equation}
which becomes
\begin{equation}
    F_\mathrm{PFA} = - \frac{\pi \rho_\mathrm{sph} \rho_\mathrm{plt}}{m^4 M^2} \left(mR - 1 + mR e^{-2mR} + e^{-2mR} \right) \left(1 - e^{- m W_1} \right) e^{- m D}~.
    \label{Ricardo-PFA}
\end{equation}
In the limit $m R \ll 1$, this gives
\begin{equation}
    F \approx \frac{2 \pi \rho_\mathrm{sph} \rho_\mathrm{plt}}{3 m^4 M^2} (m R)^3 \left( 1 - e^{- m W_1} \right) e^{- m D}~,
\end{equation}
which is in precise agreement with our screened sphere result in the appropriate limit.

As a result, we find that the proximity approximation and the small sphere result coincide as long as $R\ll 1/m$. As the proximity approximation is exact for Yukawa field theories, we find that the screening factor approximation is also exact for small spheres. We will see that for non-linear theories with screening the situation changes drastically.

\section{A screened model: The inverse power law chameleon}
%
%
%
\label{sec:cham}
As a typical example of  chameleon screening, we focus on the  self-interaction potential $V(\phi)$ and a coupling $A(\phi)$
\begin{equation}
    V(\phi) = \Lambda^4 \left(1 + \frac{\Lambda^n}{\phi^n} \right)~, \quad\quad\quad\quad A(\phi) = e^{\phi/M}~
\end{equation}
originally proposed in \cite{Khoury:2003aq} and referred to as the inverse power law chameleon model. 
In the presence of matter of density $\rho_\mathrm{m}$, the field does not respond only to the potential $V(\phi)$ but to the ``effective potential'' 
\begin{equation}
    V_\mathrm{eff} = \Lambda^4 \left(1 + \frac{\Lambda^n}{\phi^n} \right) + \frac{\phi}{M} \rho_\mathrm{m}~,
    \label{Veff-chameleon}
\end{equation}
with a corresponding Klein-Gordon equation
\begin{equation}
    \Box \phi = \frac{- n \Lambda^{4 + n}}{\phi^{n+1}} + \frac{\rho_\mathrm{m}}{M}~.
    \label{cham-eom}
\end{equation}
\subsection{The force on a small screened sphere}
Consider the case of an infinite plate, where $\rho = \infty$ for $x < 0$ and $\rho = 0$ for $x > 0$.  We therefore have $\phi(x = 0) = 0$ as a boundary condition.  It is easy to see that the solution outside the plate is
\begin{equation}
    \phi(x > 0) = \left( \frac{1}{2}\Lambda^{4 + n} (n + 2)^2 \right)^{1/(n + 2)} x^{2/(n+2)}~.
    \label{phi-plane}
\end{equation}
The screening factor of a sphere, when the ambient field value is $\phi_\mathrm{bg}$, is~\cite{Elder:2016yxm}
\begin{align} \nonumber
    \lambda &= \frac{3 M \phi_\mathrm{bg}}{\rho_\mathrm{obj} R^2}~, \\
    &= \frac{4 \pi M R \phi_\mathrm{bg}}{M_\mathrm{obj}}~.
    \label{SF-chameleon}
\end{align}
The force on a small, screened sphere is therefore
\begin{equation}
    F = - \frac{\lambda M_\mathrm{obj}}{M} \vec \nabla \phi,
\end{equation}
where $\phi$ is given by the planar solution above.  There is some ambiguity in how to pick $\phi_\mathrm{bg}$.  In our idealized case, where we are ignoring the details of the experimental setup such as the vacuum chamber walls and the backreaction of the field due to the sphere, it is given by the above planar solution.  Putting everything together, we find that the force on a small screened sphere is
\begin{equation}
    \vec F(\vec x) = - 4 \pi R \phi \vec \nabla \phi~,
    \label{force-small-cham}
\end{equation}
where $\phi(\vec x)$ is given by Eq.~\eqref{phi-plane}.  Writing this out explicitly, we have
\begin{equation}
    F = - \Lambda^2 (\Lambda R)^{4 / (2 + n)} \pi 2^\frac{4 + 3n}{2 + n} \left( (2 + n)(1 + D/R) \right)^\frac{2 - n}{2 + n}~,
    \label{force-screened-sphere}
\end{equation}
where $R$ is the radius of the sphere and $D$ is the distance between the surface of the sphere and the plate, so $x = D + R$~.

It is somewhat curious that $M$ has dropped out entirely from this expression.  This is due to the fact that we took $\phi(0) = 0$ at the surface of the plate.  If we had been more meticulous, we would have chosen a finite density for the plate, and then taken $\phi(0) = \phi_\mathrm{min}$, where $\phi_\mathrm{min}$ minimizes the effective potential.  In the limit $\rho \to \infty$, we would find $\phi_\mathrm{min} \to 0$.

Let us make clear when this approach is valid.  A rigorous derivation of the screening factor relies on several assumptions, which are summarized in \cite{Hui:2009kc}.
One of the key steps in that derivation is to draw an enclosing surface around the object of interest, which in our case is the sphere.  This surface is taken, for simplicity, to be a sphere of radius $r > R$.  On that surface, it must be possible to perform an object-background split for the field value:
\begin{equation}
    \phi \approx \phi_0(\vec x) + \phi_1(r)~,
\end{equation}
where $\phi_0$ is the environmental field (in our case, sourced by the plate) and $\phi_1$ is the field of the object in isolation.  We require that the environmental field $\phi_0$ be approximately linear throughout the region enclosed by the surface of radius $r$.  Concretely, we require
\begin{equation}
    \phi_0(x + r) \approx \phi_0(x) + \phi_0'(x)r~,
    \label{linear-background}
\end{equation}
where $x = D + R$ is the location of the center of the sphere. 

A simple way to estimate whether this is valid is to check that the first neglected term in the Taylor expansion is much smaller than the linear term, that is,
\begin{equation}
    \frac{\phi_0'(x)}{\phi_0''(x)r} \gg 1~.
    \label{screening-validity}
\end{equation}
The most stringent  condition is obtained for $r = R$.  Substituting this, and the plate solution Eq.~\eqref{phi-plane} into the above expression, we find the condition
\begin{equation}
    \left(1 + \frac 2  n \right) \left(1 + \frac D R \right) \gg 1~.
\end{equation}
This should be viewed as the minimal, necessary (but not always sufficient) condition in order for the point-sphere approximation of this section to be valid.  Let us make a few observations about this expression:
\begin{itemize}
    \item This condition is easily satisfied for $D/R \gg 1$.
    \item This is also satisfied for any $D/R$ in the limit $|n| \ll 1$.  In this limit, the evolution of the planar solution Eq.~\eqref{phi-plane} is approximately linear, so the approximation of Eq.~\eqref{linear-background} is a very good one on any scale.
    \item Notice the absence of any condition on the total size of the sphere $\Lambda R$ or its separation $\Lambda D$.  
\end{itemize}
These points make clear the advantages and limitations of this description of the force and the SFA.


\subsection{The force on a large screened sphere}
In this subsection we will compute the force on the sphere using the PFA, which should be valid when the sphere is sufficiently large.  We will explore the transition in the validity of these two regimes later on.

The first step of this approximation is to compute the pressure between two parallel plates.  Each plate is taken to be of width W, with a gap between them of width $L$ centered about $x = 0$.  The external solution was already derived in the previous section:
\begin{equation}
    \phi_\mathrm{ext}(x > L/2 + W) = \left( \frac{1}{2}\Lambda^{4 + n} (n + 2)^2 \right)^{1/(n + 2)} (x - L/2 - W)^{2/(n+2)}~,
    \label{phi-interior}
\end{equation}
and is symmetric about $x = 0$.

Next we must solve Eq.~\eqref{cham-eom} between the plates, for $|x| < L/2$.  Our boundary conditions are $\phi(\pm L/2) = 0$ and $\phi'(0) = 0$.  As before we will assume $\rho = 0$ outside the plates.

Our approach closely follows that of \cite{Brax:2007vm, Ivanov:2012cb}. Using the identity $\phi'' = \frac 1 2 \frac{d}{d \phi} (\phi'^2)$~, we can integrate Eq.~\eqref{cham-eom} from $\phi(x = 0)$ to $\phi(x < L/2)$, and we obtain
\begin{equation}
    \phi'(x) = \pm \sqrt{2\left( V(x) - V(0) \right)}
    \label{dphi-potential}
\end{equation}
We will integrate this expression again from 0 to $x$, and we find
\begin{equation}
    \pm \sqrt 2 x = \int_{\phi_0}^{\phi(x)}\frac{d \phi}{\sqrt{V(x) - V(0)}}~,
\end{equation}
where we have defined $\phi_0 \equiv \phi(0)$.
Substituting the chameleon's self-interaction potential, we find
\begin{equation}
    \pm \sqrt{2 \Lambda^{4+n}} x = \int_{\phi_0}^{\phi(x)}\frac{\phi^{n/2} d \phi}{ \sqrt{1 - (\phi / \phi_0)^n}}~.
\end{equation}
The integral on the RHS can be made more palatable by swapping variables: first to $u = \phi / \phi_0$, and then to $t = \sqrt{1 - u^n}$.  We end up with
\begin{equation}
    \pm \sqrt{\frac{n^2 \Lambda^{4 + n}}{2 \phi_0^{n + 2}}}x = \int_0^{t(x)} (1 - t^2)^{(2 - n)/2n} dt~.
\end{equation}
This integral is a particular example of the incomplete Beta function, which is defined as 
\begin{align} \nonumber
    B(z; a, b) &= \int_0^z u^{a - 1}(1 - u)^{b - 1} du.
\end{align}
As such, we find
\begin{equation}
\pm \sqrt{\frac{2 n^2 \Lambda^{4 + n}}{\phi_0^{n + 2}}}x = B \left( 1 - (\phi(x) / \phi_0)^n; \frac{1}{2}, \frac{2 + n}{2 n} \right)
\label{phi-int-implicit}
\end{equation}
Evaluating this expression at $x = L/2$, where $\phi(L/2) = 0$, gives us the central field value
\begin{equation}
    \phi_0 =  \Lambda \left( \frac{\pm n \Lambda L}{\sqrt 2 B ( \frac{1}{2}, \frac{2 + n}{2n})}  \right)^{2 / (2 + n)}~,
\end{equation}
where we have yet to determine the $\pm$ root and $B(a, b)$ is the Beta function.
Inverting Eq.~\eqref{phi-int-implicit}, we find
\begin{equation}
    \phi(x) = \phi_0 \left(1 - B^{-1} \left( \sqrt{\frac{2 n^2 \Lambda^{4 + n}}{\phi_0^{n + 2}}}x; \frac{1}{2}, \frac{2 + n}{2n} \right) \right)^{1 / n}~,
\end{equation}
where $B^{-1}$ is the inverse of the incomplete Beta function. 

Let us now compute the force.  As shown in \cite{Elder:2019yyp}  the force is
\begin{align} \nonumber
    \frac{F}{A} &= T_{x x}(x) \Big|_{L/2 + W + \epsilon}^{L/2 - \epsilon}~. \\
    &= \frac{1}{2} \phi'(L/2)^2 - V(L/2) - \frac{1}{2} \phi'(L/2 + W)^2 + V(L/2 + W)~.
\end{align}
In Eq.~\eqref{dphi-potential} we showed a way to rewrite the interior field gradient in terms of the potential.  We can do the same thing with the exterior field gradient, by integrating from $\phi(\infty)$ to $\phi(x > L/2 + W)$.  We find
\begin{equation}
    \phi'(x > L/2 + W)^2 = 2 \left( V(x) - V(\infty) \right)~,
\end{equation}
where we have assumed that the field gradient vanishes at infinity, which is justified by the explicit solution Eq.~\eqref{phi-interior}.
Combining these expressions, we find that the exact force on the plate is simply
\begin{equation}
    \frac{F}{A} = - \left( V(0) - V(\infty) \right)~.
\end{equation}
For our particular chameleon potential, this implies
\begin{equation}
    F/A = - \Lambda^4 \left( \frac{ \pm \sqrt 2 B(\frac 1 2, \frac{2 + n}{2n})}{n \Lambda L} \right)^{2n/(2+n)}~.
    \label{chameleon-plate-pressure}
\end{equation}
This agrees with the results of \cite{Brax:2007vm,Ivanov:2012cb}.

With the pressure between two parallel plates established, we can move on to the computation of the force between a plate and sphere.  We break the calculation up into infinitesimal rings, each of constant distance $L$ from the plate.  The contribution from each ring to the total force is
\begin{equation}
    dF = 2 \pi a R d\theta P(L) \cos \theta~,
\end{equation}
where $a$ is the radius of the ring, $L$ is the distance from the ring to the plate, and $P$ is the pressure given by Eq.~\eqref{chameleon-plate-pressure}.  Note that we have included an additional factor of $\cos \theta$, which allows us to include only the component of the force normal to the plate, as the other components cancel out by symmetry when integrated around the ring.

We integrate over the surface of the sphere, that is, from from polar angle $\theta = 0 \to \pi / 2$~.  Then we have $a = R \sin \theta$, $L = D + R - R\cos \theta$.
The force is then
\begin{equation}
    F = - \Lambda^4 \left( \frac{ \pm \sqrt 2 B(\frac 1 2, \frac{2 + n}{2n})}{n \Lambda R} \right)^{2n/(2+n)} 2 \pi R^2 \int_0^{\pi/2} \frac{\sin \theta \cos \theta}{(1 + D/R - \cos \theta)^{2n/(2 + n)}} d \theta~.
\end{equation}
The remaining difficulty is the non-dimensional integral
\begin{equation}
    I \equiv \int_0^{\pi/2} \frac{\sin \theta \cos \theta}{(a - \cos \theta)^b} d\theta~,
\end{equation}
with $a = 1 + D/R$.
By defining $u = \cos \theta$, this becomes
\begin{equation}
    I = \int_0^1 \frac{u }{(a - u)^b} d u~.
\end{equation}
We make a further substitution $y \equiv a - u$, so we have
\begin{equation}
    I = \int_{a - 1}^a (a y^{-b} - y^{1 - b}) dy~.
\end{equation}
Note that $b \equiv \frac{2 n}{2 + n}$, so as long as $n > 0$, we will have $b \in (0, 2)$~.  This expression then readily integrates to
\begin{equation}
    I = \begin{cases}
    \frac{1}{(1 - b)(2 - b)} \left( a^{2 - b} - (a + 1 - b)(a - 1)^{1 - b} \right) & \mathrm{if} ~ b \neq 1~, \\
    a \log \frac{a}{a - 1} - 1 & \mathrm{if} ~ b = 1~.
    \end{cases}
\end{equation}
With this, the expression for the force becomes
\begin{equation}
    F = - \Lambda^2 (\Lambda R)^{4 / (2 + n)} 2 \pi \left( \frac{\sqrt 2}{n} B\left(\frac 1 2, \frac{2 + n}{2n}\right) \right)^{2n/(2+n)}  \times I~,
    \label{force-PFA}
\end{equation}
where
\begin{equation}
    I = \begin{cases} \frac{(2 + n)^2}{4(2 - n)} \left( (1 + D/R)^{4/(2+n)} - (D/R + \frac{4}{2 + n})(D/R)^{(2-n)/(2 + n)} \right) & \mathrm{if} ~ n \neq 2~, \\
    (1 + D/R) \log(1 + R/D) - 1 & \mathrm{if} ~ n = 2~.
    \end{cases}
\end{equation}
We will use this result to extract scaling laws for the force.

\subsection{Scaling of the force with $R$}
Our expressions for the force on a small sphere (Eq.~\ref{force-screened-sphere}) and a large sphere (Eq.~\ref{force-PFA}) depend on three quantities: $\Lambda, R,$ and $D$.  But the forms of these expressions suggests a better parametrization: $\Lambda, (\Lambda R),$ and $D/R$.  The scaling of the force with $D/R$ is complicated in the case of a large sphere, as it depends on the details of the factor $I$ in Eq.~\eqref{force-PFA}.
However, for fixed $D/R$ the scaling of the force with the remaining two quantities $\Lambda$ and $\Lambda R$ is identical in each of these limiting cases:
\begin{equation}
    F \sim \Lambda^2 (\Lambda R)^{\frac{4}{2 + n}}~.
\end{equation}
This is a very fortunate turn of events.  For comparison, the symmetron force scales with the radius in different ways, depending on whether the radius is large or or small compared to the scalar field's Compton wavelength \cite{Elder:2019yyp}.  In this situation there is no such property: the force for large spheres and small ones scales in exactly the same way with $\Lambda$ and $R$.  The only remaining challenge is to understand how the $O(1)$ constants and the dependence on $D/R$ interpolate between these two regimes, where $D/R \approx 1$.  For this, we turn next to numerical solutions.  

\section{Numerical determination of the scalar force}
\label{sec:num}
In the previous sections we established two different methods to approximate the force.  The SFA, given by Eq.~\eqref{force-screened-sphere}, neglects the backreaction of the field due to the sphere, but does include some knowledge of edge effects due to the finite size of the sphere.  On the other hand, the PFA given by Eq.~\eqref{force-PFA} does the opposite: since it is based on the two parallel plate solution we have built in some knowledge of the field backreaction, but this may be a poor model of edge effects.  Our goal in this section is two-fold: first, to evaluate the accuracy Eqs.~\eqref{force-screened-sphere} and \eqref{force-PFA}, and second to interpolate between their regimes of validity.  To this end, we turn now to numerical techniques to evaluate the field and force.

We are looking for static, cylindrically symmetric solutions of Eq.~\eqref{cham-eom}:
\begin{equation}
    \vec \nabla^2 \phi = - \frac{n \Lambda^{4+n}}{\phi^{n+ 1}} + \frac{\rho_\mathrm{m}}{M}~.
\end{equation}
We make a further simplifying assumption, consistent with the previous sections.  First, we take the screened limit, in which $\phi$ at the surface of the sphere and the plate is $\approx 0$.  Second, we neglect the density of the residual gas inside the vacuum chamber.  These assumptions are justified when $\phi_\mathrm{surface} \ll \phi_\mathrm{amb}$, where $\phi_\mathrm{surface}$ is the field value at the surface of the sphere and $\phi_\mathrm{amb}$ is the ambient field value.

With the above assumptions, $\rho_\mathrm{amb}/M$ drops out of the equation of motion, and we have the boundary condition $\phi = 0$ at the surfaces of the sphere and plate.  We are now able to eliminate $\Lambda$ from the equation of motion by the following rescaling of the coordinates and field variable:
\begin{equation}
    \hat \phi \equiv \frac{\phi}{\Lambda}~, \quad \quad \hat x \equiv \Lambda x~.
    \label{nondim-rescaling}
\end{equation}
The equation we are left to solve is
\begin{equation}
    \hat \nabla^2 \hat \phi = - \frac{n}{\hat \phi^{n + 1}}~,
    \label{nondim-eom}
\end{equation}
where $\hat \nabla^2$ is the Laplacian in the rescaled coordinates.  A given solution to the sphere-plate system is now completely defined by its rescaled parameters $\Lambda D$ and $\Lambda R$.

Examining our analytical solutions, given by Eqs.~\eqref{force-screened-sphere} and \eqref{force-PFA}, suggests that we parameterize a solution with $\Lambda R$ and $D / R$.  We expect the SFA to work well when $D / R \gg 1$, while the PFA is expected to work when $D / R \ll 1$.  The intermediate region, in which $0.1 \ll D / R \ll 10$, is unknown, but fortunately this is precisely the region in which numerical solutions are most easily obtained.

One could scan over a range of different values for $\Lambda R$ and $D / R$, repeating the process for different values of the chameleon model parameter $n$.  A full 3-dimensional parameter sweep would take significant computational resources to cover.  Fortunately, this situation is ameliorated by an additional hidden scaling symmetry.  Notice that the SFA and the PFA both scale as
\begin{equation}
    F \sim (\Lambda R)^{4 / (2 + n)}~.
    \label{R-scaling}
\end{equation}
This suggests that the numerical solutions might scale the same way, and our results will confirm this fact.

We use a commercially available PDE solver to solve Eq.~\eqref{nondim-eom} in cylindrical coordinates.  The field solution is a function of the height and radial coordinates, $\hat z$ and $\hat r$ respectively, where the hat indicates that the coordinate is nondimensional via Eq.~\eqref{nondim-rescaling}.  The solution is obtained within a simulation box $0 < \hat z < \hat D + 2\hat R + \hat H$ and $0 < \hat r < \hat R + \hat H$, where $\hat H$ is a parameter to set the overall box size.  It is chosen to be large enough that the computed value for the force is independent of this choice.

\begin{figure}[t]
\centering
\includegraphics[width=0.75\textwidth]{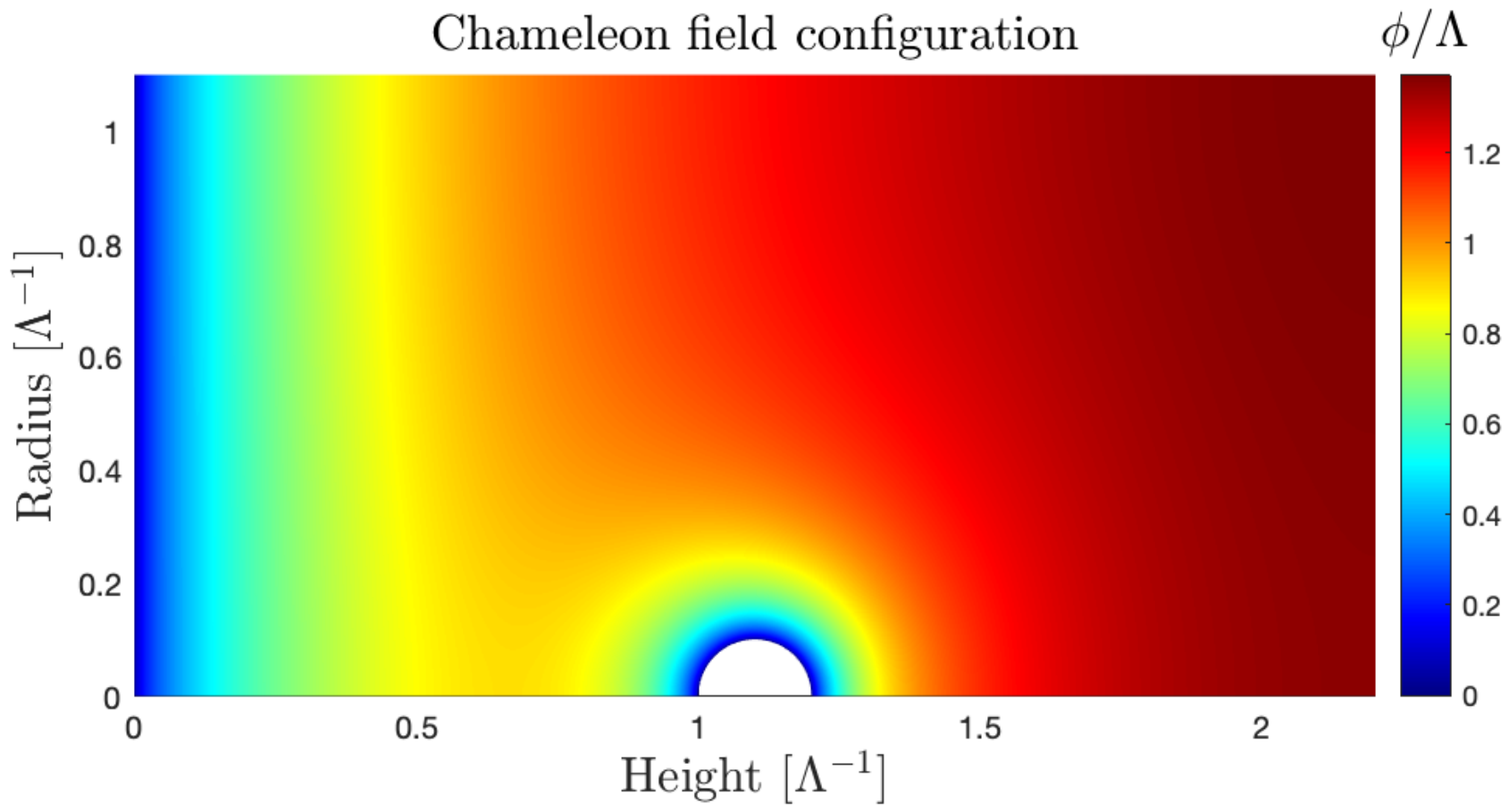}
\caption{\small Sample numerical field solution.  The boundary conditions $\phi = 0$ at the left side and on the ball in the middle represent the plate and the sphere.   Solution parameters were $\Lambda D = 1, ~\Lambda R = 0.1, ~\hat H = 1$.  As expected, the field rolls to a large value far away from the sphere and plate, and is suppressed between the two objects.}
\label{fig:sample-solution}
\end{figure}

We set $\hat \phi = 0$ on the edge where $\hat z = 0$, which simulates the flat plate.  Likewise, we set $\hat \phi = 0$ on a ball of radius $\hat R$ centered at $\hat z = \hat D + \hat R, \hat r = 0$, which naturally represents the sphere.  On all other boundaries we enforce the condition $\hat \nabla \hat \phi = 0$ normal to the boundary.  This is appropriate for the boundaries on $\hat r = 0$, but is clearly unphysical for the other boundaries.  However, the ill effects of this unphysicality are avoided by choosing $\hat H$ to be sufficiently large.  This setup is illustrated by the sample numerical field solution in Fig.~\ref{fig:sample-solution}.

Once a numerical solution is obtained, the field may be used to calculate the chameleon sphere-plate force.  Our method is the same one used in a previous study of the symmetron~\cite{Elder:2019yyp}, which is based on the Einstein-Infeld-Hoffman method to compute the force on extended objects in general relativity~\cite{Hui:2009kc,Einstein:1938yz}.  We review the main points here.  First, note that the total momentum in a spatial volume $\cal V$ is
\begin{equation}
    P_i = \int_{\cal V} d^3 x T_i^0~,
\end{equation}
where $T_{\mu \nu}$ is the energy-momentum tensor for all scalar and matter fields within the volume.  The time derivative of the momentum is the force, and may be simplified as follows:
\begin{equation}
    \dot P_i = \int_{\cal V} d^3 x \partial_0 T_i^0 = - \int_{\cal V} d^3 x \partial_j T_i^j = - \int_{\cal B} d^2 \sigma_j T_i^j~,
\end{equation}
where we have used conservation of the energy-momentum tensor as well as the divergence theorem to turn the volume integral into a surface integral over the volume's boundary $\cal B$.  The vector for the infinitesimal area $d^2 \sigma_j$ is normal to the boundary.  We will compute this surface integral for a volume centered on the sphere.  If this boundary were set exactly at the surface of the sphere the evaluation of this integral would require knowledge of the matter fields.  However, by increasing the size of the volume slightly, we guarantee that the matter fields are zero and the energy-momentum tensor is given only by that of the scalar field:
\begin{equation}
    T_\mathrm{\mu \nu} = \partial_\mu \phi \partial_\nu \phi + \eta_{\mu \nu} \left( - \frac{1}{2} (\partial \phi)^2 - V(\phi) \right)~.
\end{equation}

\begin{figure}[t]
\centering
\begin{subfigure}{0.45\textwidth}
    \centering
    \includegraphics[height=2.25in]{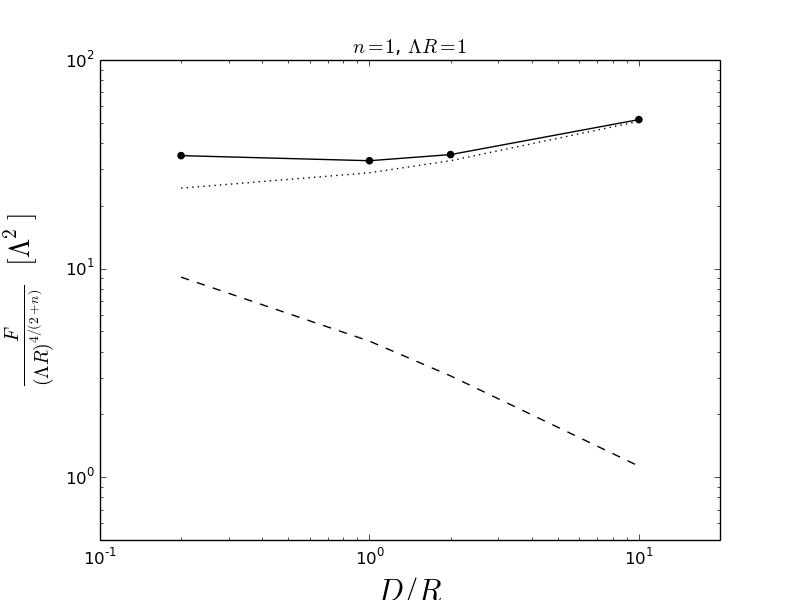}
    \caption{}
    \label{fig:numerics-a}
\end{subfigure}
\begin{subfigure}{0.45\textwidth}
    \centering
    \includegraphics[height=2.25in]{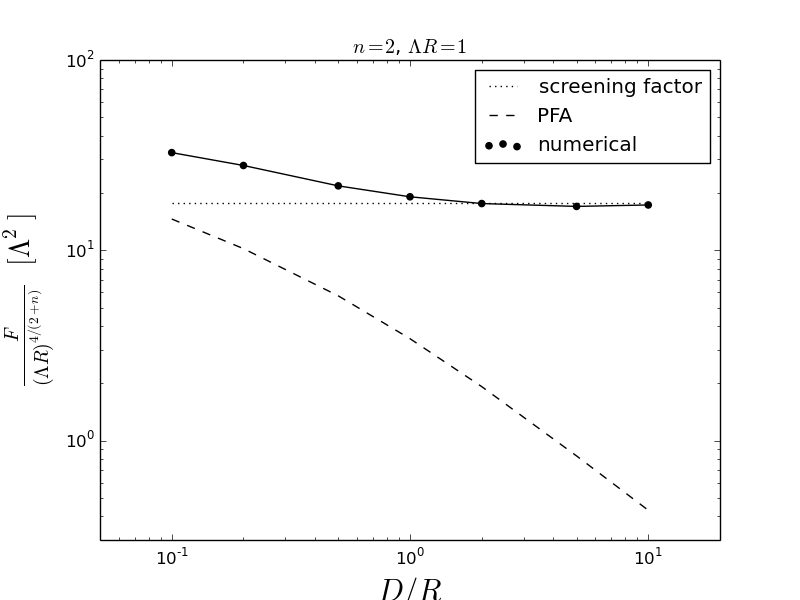}
    \caption{}
    \label{fig:numerics-b}
\end{subfigure}
\begin{subfigure}{0.45\textwidth}
    \centering
    \includegraphics[height=2.25in]{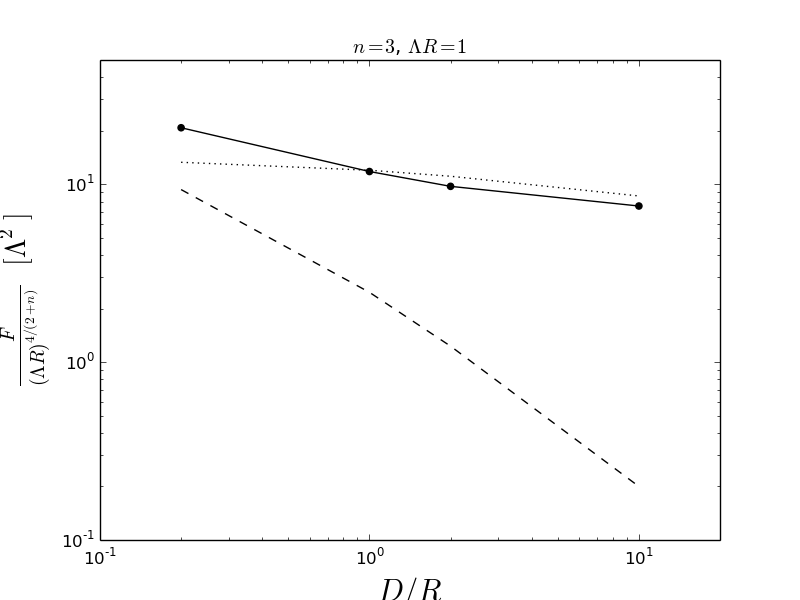}
    \caption{}
\end{subfigure}
\begin{subfigure}{0.45\textwidth}
    \centering
    \includegraphics[height=2.25in]{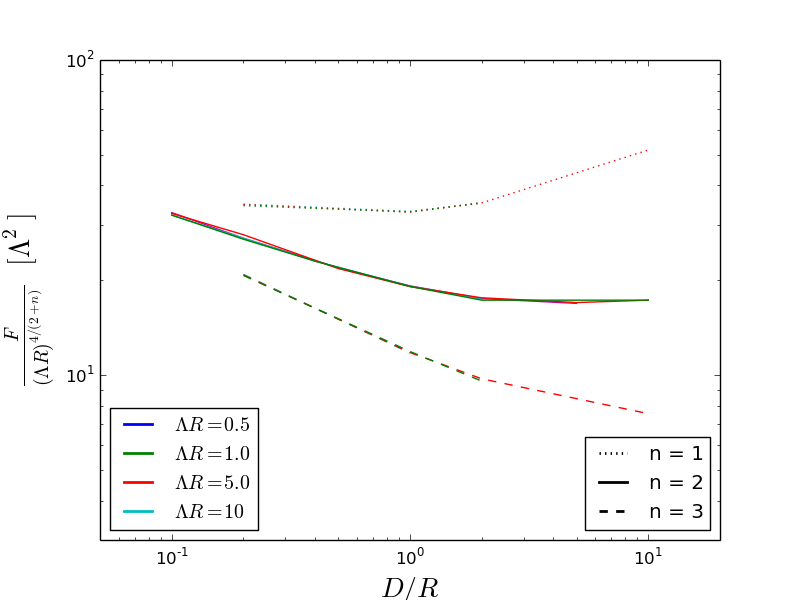}
    \caption{}
    \label{fig:numericsd}
\end{subfigure}
 \caption{\small Comparison of numerical data to the PFA and SFA for the $n = 1, 2,$ and $3$ chameleon.  The scaling with $\Lambda R$ is shown to be highly regular, as predicted (plot d).  We see that for realistic Casimir experiments where $D/R \sim 1$, the screening factor is a more accurate approximation method.}
 \label{fig:numerics}
\end{figure}

This procedure was done for $0.1 \leq \Lambda R \leq 10$, $0.1 < D/ R < 10$, and $n = 1, 2, 3$, and our results are shown in Fig.~\ref{fig:numerics}.  First, we note that Fig.~\ref{fig:numericsd} supports our hypothesis that the force scales as Eq.~\eqref{R-scaling}, where we have checked for $n = 1, 2, ~\mathrm{and}~ 3$.  Consequently, we are free to set $\Lambda R = 1$, knowing that we can use Eq.~\eqref{R-scaling} to scale the result as needed.  This allows us to focus our attention and computational resources on the remaining two parameters $D/R$ and $n$.

For all three values of $n$, we see good agreement with the screening factor approximation when $D / R \approx 10$.  In fact, the screening factor approximation does surprisingly well throughout the entire region that was tested, differing at most by $\sim 30\%$ from the exact numerical solution.  On the other hand, the proximity force approximation does a surprisingly poor job, even in the regime where it was expected to work well, $D/R \approx 0.1$.  Throughout the entire window that was tested numerically, the screening factor is in fact a better predictor of our exact solutions.  It is entirely possible that the proximity force approximation works well for even more extreme values $D/R \ll 0.1$, but we are unable to test this region numerically.  What we have found, however, is that the proximity force approximation is an inappropriate tool in the analysis of practical Casimir experiments using a plate and a sphere, as those experiments typically have $D/R \gtrsim 0.1$.  As such, when we place experimental bounds on the chameleon theory, we will employ the screening factor approximation, not the proximity force approximation.

Finally, we briefly comment on the curious-looking behavior shown by the SFA in Figs.~\ref{fig:numerics-a} and \ref{fig:numerics-b}.  Namely, that the screening factor approximation, as well as our numerical results, show the chameleon force being constant, or even increasing, with increasing distance from the plate.  Intuitively, one might expect that the force should decrease with distance, although this need not be the case in the highly idealised situation we consider here.  We also remind the reader that in standard electrostatics, an infinitely large charged plate also sources a force that is constant with distance.  The resolution in this case is that we have neglected the density of the residual gas $\rho_\mathrm{vac}$ surrounding the plate and sphere, as well as the finite size of the plate.  At a distance of order a Compton wavelength $m_\mathrm{eff}^{-1}$, which will be discussed in further detail in the next section, the force law will become exponentially suppressed.

\section{Constraints on chameleon modified gravity}
\label{sec:exp}
In this section we describe a realistic experiment, based on the current state of the art, that could be performed in the near future.  The experimental parameters are drawn from \cite{Chen:2014oda, Bimonte:2021sib, Elder:2019yyp} and are as follows.  A sphere of radius $R = 150 ~\mathrm{\mu m}$ is suspended $15~\mathrm{\mu m}$ away from a rotating disk.  The disk has $50 \mathrm{\mu m}$ deep trenches spaced periodically, such that as the disk rotates the distance between the disk and the nearest point of the sphere varies between $D_\mathrm{near} = 15~\mathrm{\mu m}$ and $D_\mathrm{far} = 65~\mathrm{\mu m}$.  The force between the sphere and the disk is monitored, and the difference in the force between the $D_\mathrm{near}$ and $D_\mathrm{far}$ configurations is measured to be $\delta F = F_\mathrm{near} - F_\mathrm{far} < 0 \pm 0.2~\mathrm{fN}$.  Note that the Casimir force itself is much smaller than this beyond a distance of $\approx 10\mu\mathrm{m}$, eliminating the need to  accurately model the Casimir force in order to derive constraints~\cite{Chen:2014oda}.

These experimental parameters may be used to constrain fundamental theories, as any new scalar must contribute a force that is smaller than $\delta F$.  However, we emphasize that these are only forecasts for a near-future hypothetical experiment.  The experiment that these values are drawn from used a plate with trenches that were only $200~\mu\mathrm{m}$ apart.  This both suppresses the chameleon force in the measurement and also necessitates accurate modeling of the effects from the trench walls, which is beyond the scope of this work.  As such, we must content ourselves with placing forecasts on a future experiment, with similar parameters, that uses much wider trenches to avoid these issues.

In this experiment, the ratio $D/R$ varies between 0.1 and 0.43.  We can see from Fig.~\ref{fig:numerics-a} that in this regime the screening factor approximation is perfectly adequate to estimate the force on the sphere, and is considerably more straightforward to employ than the proximity force approximation or the numerical solutions.  As such, in this section we will exclusively employ the screening factor approximation, given by Eq.~\eqref{force-screened-sphere}, to place bounds on the $n = 1$ chameleon.  The differential force is
\begin{equation}
    \delta F = \Lambda^2 (\Lambda R)^{4 / (2 + n)} \pi 2^\frac{4 + 3n}{2+n} (2 + n)^\frac{2 - n}{2+n} \left( \left(1 + D_\mathrm{near}/R \right)^\frac{2 - n}{2+n} - \left(1 + D_\mathrm{far}/R \right)^\frac{2 - n}{2+n} \right)~.
    \label{SF-dF}
\end{equation}

\begin{figure}[t]
    \centering
    \includegraphics[width=0.75\textwidth]{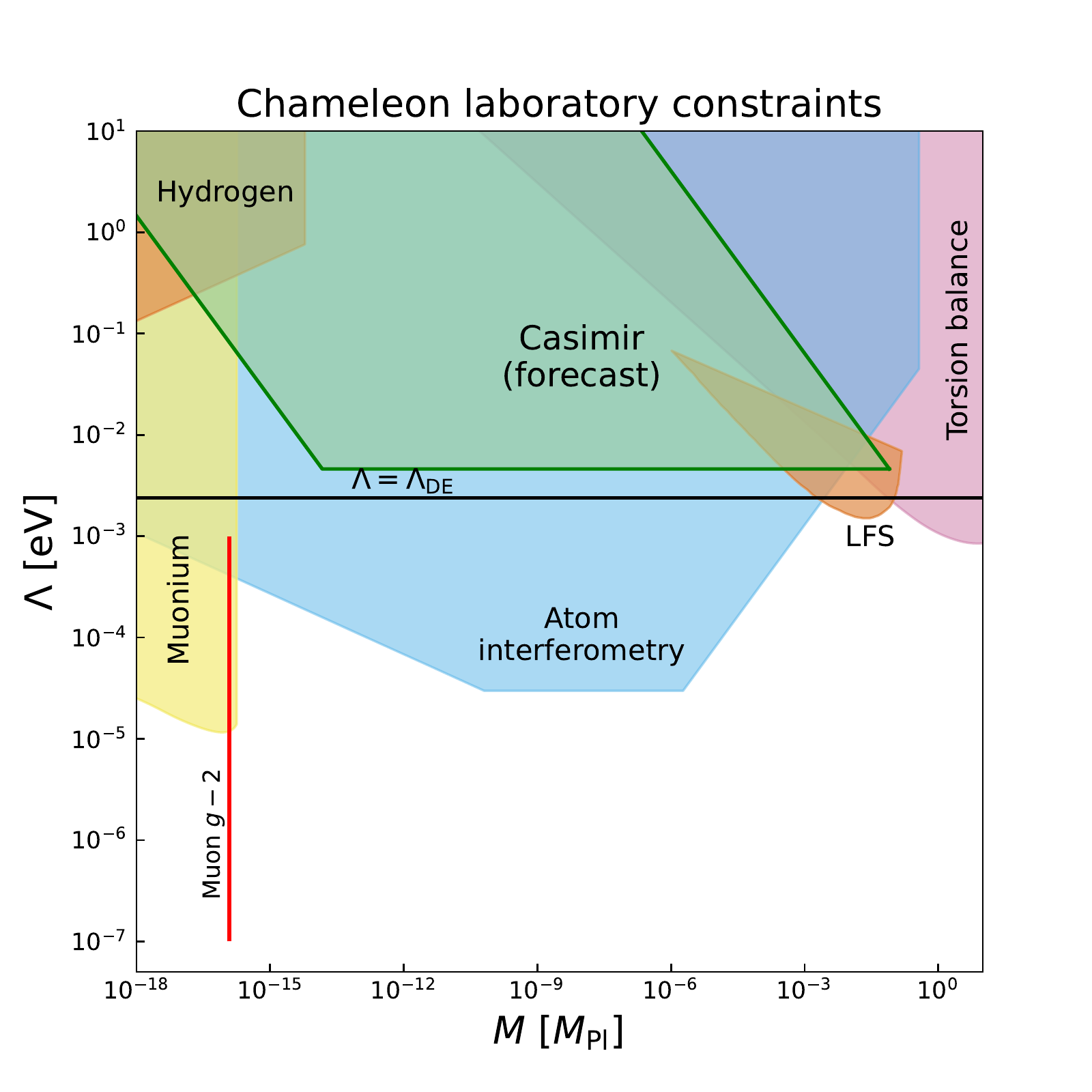}
    \caption{Prospective bounds on the parameters for an $n = 1$ chameleon (left) vs. combined constraints from various experiments.  It is clear that Casimir experiments are sensitive to a wide range in $M$, and accesses the small window that has also recently been tested by a levitated force sensor, marked as ``LFS''~\cite{2022NatPh..18.1181Y}.}
    \label{fig:bounds}
\end{figure}

\begin{figure}[t]
\centering
\begin{subfigure}{0.45\textwidth}
    \centering
    \includegraphics[height=2.25in]{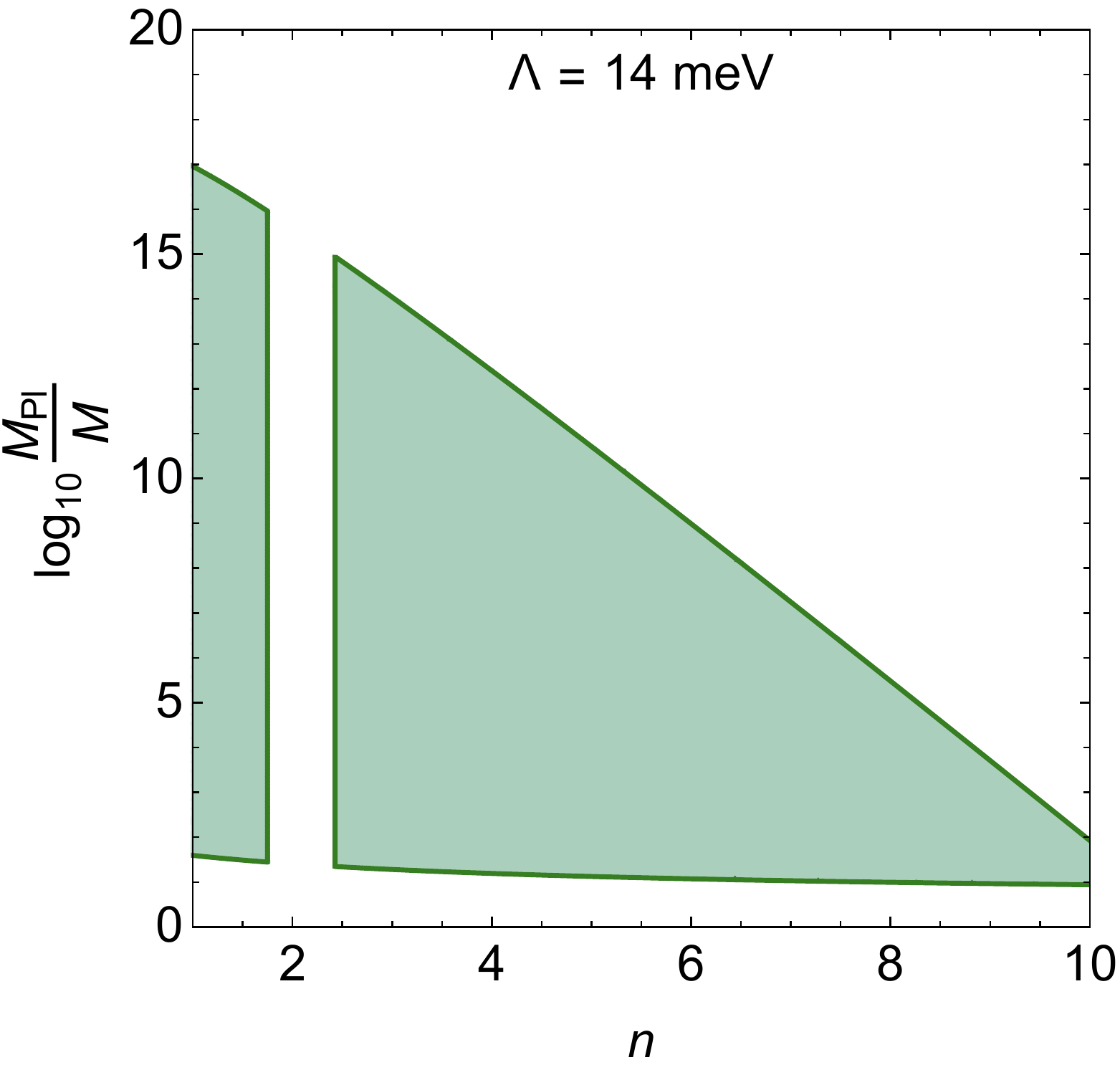}
    \caption{}
    \label{fig:beta-n-a}
\end{subfigure}
\begin{subfigure}{0.45\textwidth}
    \centering
    \includegraphics[height=2.25in]{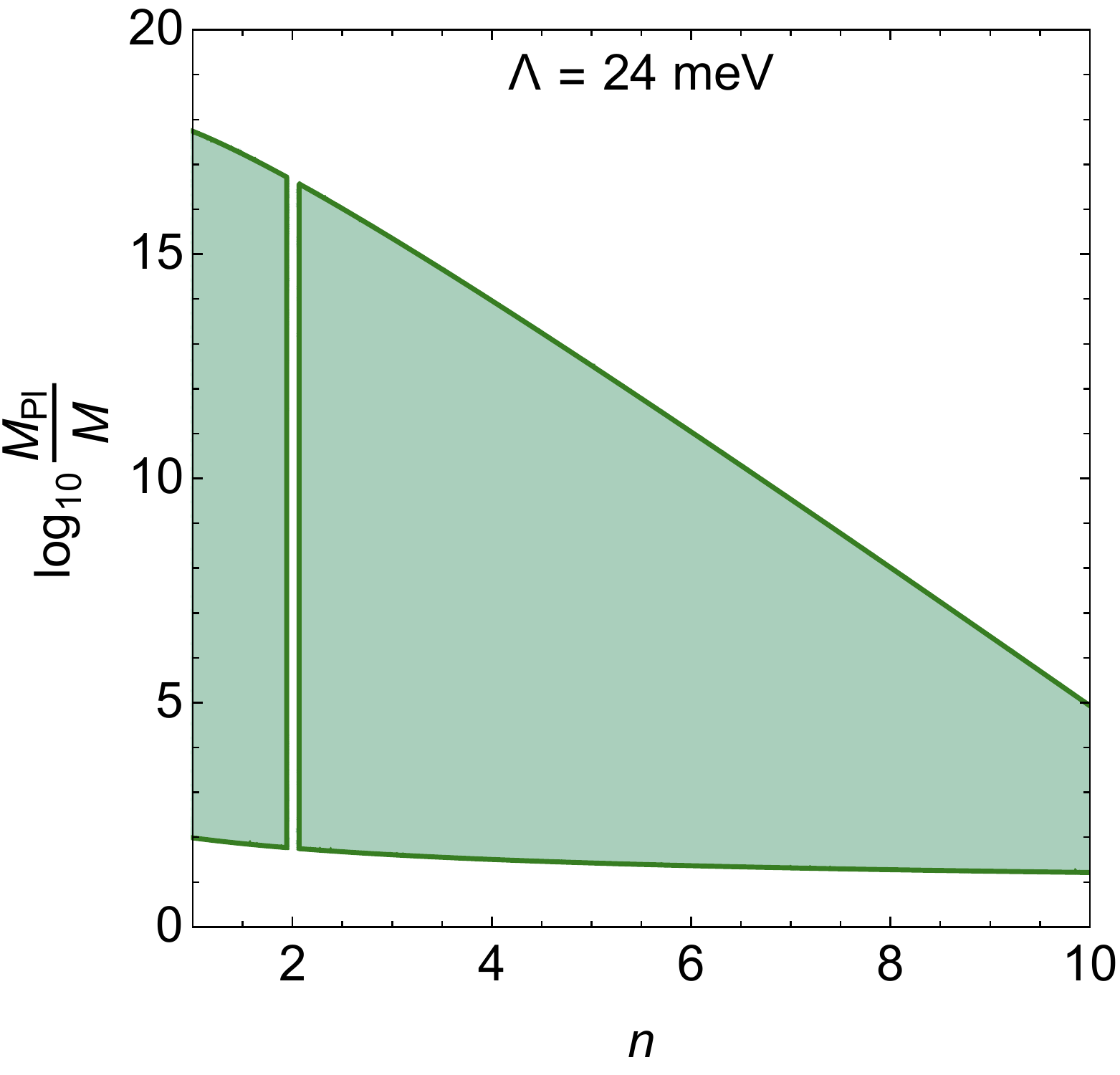}
    \caption{}
    \label{fig:beta-n-b}
\end{subfigure}
 \caption{\small Forecast constraints vs. chameleon parameter $n$, at particular values of $\Lambda$.  These constraints are made using the screening factor approximation, which predicts zero differential force in the vicinity of $n = 2$, hence the gaps in that region.}
 \label{fig:beta-n}
\end{figure}

There are several limitations to this expression that are worth noting.  First, it was assumed that the sphere is screened.  Using Eq.~\eqref{SF-chameleon}, this amounts to demanding
\begin{equation}
    \frac{3 M \phi_\mathrm{bg}}{\rho_\mathrm{obj} R^2} < 1~,
    \label{ineq-screened}
\end{equation}
where the ambient field value $\phi_\mathrm{bg}$ in the vicinity of sphere is given by evaluating the planar solution Eq.~\eqref{phi-plane} at $x = D_\mathrm{far} + R$.  Second, we have neglected the finite density of the residual gas inside the vacuum chamber of the experiment.  This is accurate provided that the distance to the plate is shorter than the scalar field's Compton wavelength inside the vacuum chamber
\begin{equation}
    (D_\mathrm{far} + R) < m_\mathrm{eff}^{-1}~.
    \label{ineq-yukawa}
\end{equation}
The Compton wavelength is the inverse of the mass of the scalar perturbations, and is computed as follows.  The mass of the perturbations around a particular field value is
\begin{equation}
    m_\mathrm{eff}^2 = \frac{d^2}{d \phi^2} V_\mathrm{eff}(\phi)~,
\end{equation}
where the Chameleon effective potential $V_\mathrm{eff}$ is given by Eq.~\eqref{Veff-chameleon}.  This is to be evaluated about the minimum of the effective potential, which is set by the ambient density inside the vacuum chamber.  For this experiment, we have $\rho_\mathrm{amb} = 10^{-13} ~\mathrm{g / cm}^3$.  For a chameleon potential, the scalar mass is
\begin{align} \nonumber
    m_\mathrm{eff}^2(\phi_\mathrm{min}) &= n (n + 1) \frac{\Lambda^{4+n}}{\phi^{n + 2}_\mathrm{min}}~, \\
    \phi_\mathrm{min} &= \left( \frac{n \Lambda^{4+n} M}{\rho_\mathrm{amb}} \right)^{1 / (n+1)}~.
\end{align}
The region where Eqs.~\eqref{ineq-screened},~\eqref{ineq-yukawa} are satisfied, and where $\delta F \leq 0.2 ~\mathrm{f N}$, are plotted in Figs.~\ref{fig:bounds}~and~\ref{fig:beta-n}.  In Fig.~\ref{fig:bounds} we see that the forecast is tantalizingly close to ruling out $n = 1$ chameleons at the dark energy scale. An order of magnitude improvement in the sensitivity of Casimir-type experiments would close this gap and conclusively rule out those models.
This gap has recently been tested using a levitated force sensor~\cite{2022NatPh..18.1181Y} and therefore a Casimir experiment would corroborate those findings and could even put the constraints in region of the parameter space so far  only probed by atomic interferometry. This would be an important method of confirming the exclusion regions already known in the literature. In Fig.~\ref{fig:beta-n} we show the dependence of the constraints on the chameleon index $n$.  Apart from the spurious gap at $n = 2$, which is due only to our use of the screening factor approximation, we see significant constraining power across a wide range of $n$ and $M$ values.

Present forecasts do not show an ability to quite reach the dark energy scale $\Lambda = 2.4~\mathrm{meV}$, although it is possible that mild improvements in the near future could enable this.  Using Eq.~\eqref{SF-dF}, we see that, if keeping $R, D_\mathrm{near}, D_\mathrm{far}$ fixed, it will be possible to constrain down to the dark energy scale if the experimental sensitivity is improved by a factor of 40, to rule out differential forces $\delta F > 5~\mathrm{aN}$.

We now briefly comment on the $n = 2$ chameleon.  This is a special case for the screening factor approximation, which predicts $\delta F = 0$ as seen in Eq.~\eqref{SF-dF} and Fig.~\ref{fig:beta-n}.  However, we note that our numerical results, shown in Fig.~\ref{fig:numerics-b} do in fact display a variation in the force with the distance.  From the numerical data, we have
\begin{equation}
    \delta F \approx 10 \Lambda^3 R~,
\end{equation}
which rules out $\Lambda > 7~\mathrm{meV}$ within the inequalities given by Eqs.~\eqref{ineq-screened},~\eqref{ineq-yukawa}~.
Finally in Fig.~\ref{fig:beta-n} we have represented the parameter space of the chameleon models for a given value of $\Lambda$ and different coupling scales $M$ and index $n$. As can be seen, although the value of $\Lambda$ corresponding to the dark energy scale will not be reachable unless the sensitivity of future Casimir experiments is increased by an order of magnitude, larger values of $\Lambda$ would open up the parameter space. Values of $n$ up to $n=10$ and $\Lambda$ varying over nearly fifteen orders of magnitude will be testable. Future Casimir experiments will probe chameleons in uncharted regions of their parameter space.

Before concluding, we remark on potential improvements to this setup that would maximise the chameleon signal.  If one is operating within the regime that the SFA works well, then Eq.~\eqref{SF-dF} shows that three attributes are desirable: (i) $R$ as large as possible, (ii) $D_\mathrm{near}$ as small as possible, and (iii) $D_\mathrm{far}$ as large as possible, all within the limits of applicability of the SFA and practical limitations.  However, if one goes beyond the region of applicability of the SFA a richer story emerges.  Although we were unable to confirm this fact numerically, it is likely that the PFA is accurate for very small values of $D/R$.  Indeed, Fig.~\ref{fig:numerics-a} shows that the numerical determination of the force appears to increase at smaller values of $D/R$, a trend which may continue to even smaller values to match onto the PFA result.  If this is the case, then the force curve has a local minimum at $D/R \approx 1$.  In this case, the optimal setup would be to have (i) $R$ as large as possible, (ii) $D_\mathrm{near}$ as small as possible, and (iii) $D_\mathrm{far}$ situated at the minimum of this curve, at $D/R \approx 1$.  Alternatively, one could have $D_\mathrm{near}$ at the minimum and $D_\mathrm{far}$ as large as possible, although this would sacrifice sensitivity to forces with range smaller than $D_\mathrm{near}$.

\begin{figure}[t]
\centering
    \includegraphics[width=0.8\textwidth]{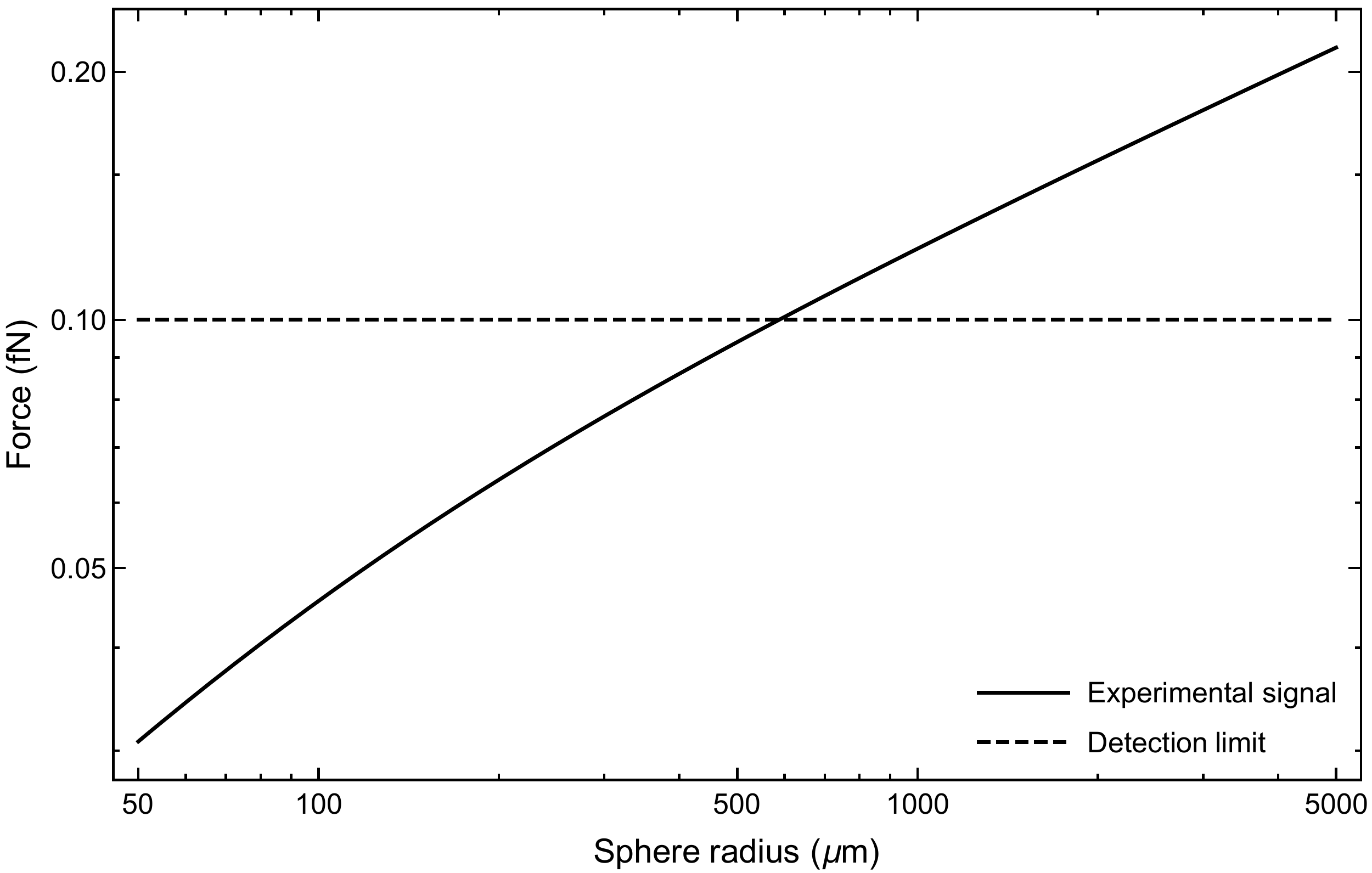}
 \caption{\small Sensitivity of experimental setup to $n = 1$ chameleons at the dark energy scale $\Lambda = 2.4~\mathrm{meV}$.  The experimental sensitivity is assumed to be $0.1~\mathrm{fN}$, the sphere is $D_\mathrm{near} = R$, and the trenches are $200~\mathrm{\mu m}$ deep so that $D_\mathrm{far} = R + 200~\mathrm{\mu m}$. 
 We see that an experiment with these parameters will be able to test chameleons at the dark energy scale provided that the sphere radius is $R \gtrsim 600~\mathrm{\mu m}$.}
 \label{fig:optimize-setup}
\end{figure}

Going slightly further, we consider another scenario which would enable detection of chameleons at the dark energy scale $\Lambda = 2.4~\mathrm{meV}$.  In this picture, The experimental sensitivity is further increased to $0.1~\mathrm{fN}$, and the trench depth is increased to $200~\mathrm{\mu m}$.  We set $D_\mathrm{near} = R$, so that we are safely within the regime in which the SFA is confirmed to be reasonably accurate (as seen in Fig.~\ref{fig:numerics-a}).  The experimental sensitivity of this setup is plotted in Fig.~\ref{fig:optimize-setup} as a function of the sphere radius.  We see that increasing the sphere radius to $R \gtrsim 600~\mathrm{\mu m}$ would be sufficient to test chameleons down the dark energy scale $\Lambda = 2.4~\mathrm{meV}$. 

\section{Conclusions}
\label{sec:conclusions}
In this paper we have compared the numerical results for non-linear field theories inducing a classical force between a plate and a sphere with the Proximity Force Approximation, commonly used to evaluate the Casimir interaction between non-planar objects. We have have focused on non-linear scalar field theories which modify gravity and are screened in dense environments. More particularly, our analysis has been performed with inverse power law chameleons. After recalling that the PFA is exact for linear theories such as Newtonian gravity or Yukawa interactions, we have shown how the PFA can be implemented for chameleons and compared it to another approximation method, the Screening Factor approximation, which is commonly used to calculate the force on extended bodies in screened theories. It turns out that for chameleons we find a good agreement between the SFA and numerical results whilst the PFA fares rather poorly, even in the regime in which it was expected to work well.  This is also the regime in which current state-of-the-art Casimir force sensors operate.

Using the SFA we can forecast what future Casimir experiments should measure in particular in the $n=1$-inverse chameleon case as well as for more general values of $n$. We have shown that future Casimir experiments will test chameleon models in a wide range of parameters for larger values of the potential scale $\Lambda$ than the dark energy scale, and for indices $n$ characterising the inverse power law of the interaction potential as large as $n=10$ for coupling scales to matter ranging over as much as fifteen orders of magnitude.  We also showed that moderate improvements in the sensitivity of these experiments will allow access to $\Lambda$ at the dark energy scale. The numerical techniques that we have used to validate the SFA compared to the PFA could also be amenable to extensions closer to exact experimental situations, such as ones with more novel geometries~\cite{bsaibes2020} and where dynamical effects could be taken into account, i.e. the case of moving sphere over step-like geometries. A treatment of these more realistic setups, both numerically and analytically, are left for future work. 

{\bf Acknowledgments}
The authors are grateful to Clare Burrage, Joerg Jaeckel, and particularly Ricardo Decca for helpful discussions.  B.E. was supported by a Leverhulme Trust Research Leadership Award while completing a portion of this work.  P.B. acknowledges support under CERN-TH-2022-191.

\renewcommand{\em}{}
\bibliographystyle{utphys}
\addcontentsline{toc}{section}{References}
\bibliography{main}

\end{document}